\documentclass{amsart}

\theoremstyle{definition}

\theoremstyle{remark}

\DeclareMathOperator*{\tr}{tr}

\begin{document}

\title[Some open problems]{Some open problems in random matrix theory and the theory of integrable systems}

\author{Percy Deift}
\address{Courant Institute of Mathematical Sciences,
         New York University,
         251 Mercer Street,
         New York, NY 10012}
\email{deift@cims.nyu.edu}
\thanks{The work of the author was supported in part by NSF grant DMS-0500923. The author would like to thank Irina Nenciu for
extensive help and suggestions in preparing this manuscript.}

%    General info
\date{November 6, 2007 and, in revised form, November 23, 2007.}



\begin{abstract}
We describe a list of open problems in random matrix theory and integrable systems which was
presented at the conference ``Integrable Systems, Random Matrices, and Applications''
at the Courant Institute in May 2006.
\end{abstract}

\maketitle

\section*{Introduction}
During the conference ``Integrable Systems, Random Matrices, and Applications,'' held
at the Courant Institute in May 2006, the organizers asked
me to present a list of unsolved problems. What follows is, more or
less, the list of problems I presented, written down in an informal style. Detailed
references are readily available on the web. In the text, various authors are mentioned by name: this is
to aid the reader in researching the problem at hand, and is not meant in any way to be a detailed, historical
account of the development of the field. I ask the reader for his'r indulgence on this score.
In addition, the list of problems is not meant to be complete or definitive: it is simply
an (unordered) collection of problems that I think are important and interesting, and which I would very much
like to see solved.

\bigskip

\textbf{Problem 1. KdV with almost periodic initial data.} Consider the Korteweg-de Vries (KdV) equation
\begin{equation}\label{E:1}
u_t+uu_x+u_{xxx}=0
\end{equation}
with initial data
\begin{equation}\label{E:2}
u(x,t=0)=u_0(x),\qquad x\in\mathbb{R}\,.
\end{equation}
In the 1970s, McKean and Trubowitz proved the remarkable result
that if the initial data $u_0$ is periodic,
$u_0(x+p)=u_0(x)$ for some $p>0$, then the solution $u(x,t)$ of \eqref{E:1} is almost periodic
in time. The conjecture is that the same result is true if $u_0(x)$ is almost periodic. This turns out to be
an extremely difficult problem: for such initial data, even short time existence for
\eqref{E:1} is not known. Loosely speaking, the difficulty centers around the fact that,
in the periodic case, various $L^2$ quantities are conserved, e.g.
$$
\int_0^p u^2(x,t)\,dx=\int_0^p u_0^2(x)\, dx,\qquad t\geq0,
$$
but in the almost periodic case, we only know that averaged $L^2$ quantities, such as
$\lim_{L\to\infty} \frac{1}{2L}\int_{-L}^L u^2(x,t) dx$, are conserved. It is not clear
how to use such averaged $L^2$ quantities to analyze the Cauchy problem for \eqref{E:1}.
Furthermore, the result of McKean and Trubowitz relies in an essential way
on the integrability of KdV: in particular, $H(t)=-\frac{d^2}{dx^2}+u(x,t)$
undergoes an isospectral deformation under \eqref{E:1}, and the periodic
spectrum of $H(t)$ provides integrals for the flow. In the almost periodic case,
assuming that a solution $u(x,t)$ exists for $t>0$, the same is true
for $\textrm{spec}(H(t))$, which equals $\textrm{spec}(H(t=0))$; the problem is that very little
is known about $\textrm{spec}(H(t=0))$ for general, almost periodic $u_0(x)$. The spectrum may have
absolute continuous, singular continuous, and pure point components. For example,
in famous work from the 1970s, Dinaburg and Sinai used KAM techniques
to show that $H=-\frac{d^2}{dx^2}+\cos x +\cos (\sqrt{2}x)$ has (some)
absolutely continuous spectrum. And in the 1980s Moser constructed an example of a limit periodic potential
$u_0(x)$ for which $H=-\frac{d^2}{dx^2}+u_0(x)$ has Cantor spectrum. And even if the spectral
theory of $-\frac{d^2}{dx^2}+u_0(x)$ for general almost periodic $u_0(x)$ was known,
that would only be the first step in understanding the short time, and then the long time,
behavior of the solution of \eqref{E:1}.

\bigskip

\textbf{Problem 2. Universality for Random Matrix Theory (RMT).} Universality for unitary matrix ensembles ($\beta=2$)
in the bulk and at the hard and soft edges is now rather well understood. For orthogonal and
symplectic ensembles ($\beta=1$ and $\beta=4$ respectively) universality in the bulk
and at the hard and soft edges is now understood in the case that the underlying
distribution is of the form $\frac{1}{Z_N} e^{-\tr Q(M)}\,dM$, where
\begin{equation}\label{E:3}
Q(x)=a_0 x^m+a_1 x^{m-1}+\cdots
\end{equation}
is a polynomial. It is of considerable interest to prove universality in the
cases $\beta=1$ and $\beta=4$ for more general potentials $Q(x)$, in particular for
$Q(x)$ of the form
\begin{equation}\label{E:4}
Q(x)=NV(x),
\end{equation}
where $V(x)$, is, say, a polynomial. In case \eqref{E:3}, the associated
equilibrium measure is supported on a single interval, whereas in case
\eqref{E:4} the equilibrium measure may be supported on a union
of disjoint intervals, and this makes the problem considerably harder.
Following Tracy-Widom and Widom, one still utilizes known asymptotics for polynomials orthogonal
with respect to the weight $e^{-NV(x)}\,dx$ on $\mathbb{R}$, but now one works on a Riemann surface
of genus $g>0$ as opposed to the Riemann sphere in case \eqref{E:3}. The real
difficulty involves a certain determinant, $D_m$ say, that arises in the analysis.
From the algebraic point of view, $D_m$ can be viewed as controlling the change of basis from orthogonal
polynomials to skew orthogonal polynomials, as the size $N$ of the polynomials goes to infinity.
From an analytical/physical
point of view, $D_m$ arises in the computation of a ratio of partition functions,
again in the large $N$ limit. The essential task here is to show that $D_m\neq0$. In case
\eqref{E:3} this follows from a lengthy ad hoc calculation; in the case \eqref{E:4}, the problem
is wide open, though recent work of Shcherbina may point the way. Alternatively, Lubinsky
has recently developed very interesting new methods for unitary ensembles. Can these methods be adapted
to prove universality for the orthogonal and symplectic ensembles?

For Wigner ensembles, universality at the edge (Soshnikov,...) is now well understood for a wide
variety of distributions on the entries of the matrices. It is a long-standing
conjecture, now more than forty years old, that universality is also true in the bulk,
but there has been very little progress. At the edge, moment
methods are powerful enough to prove universality, but in the bulk they
provide no information. By contrast, for invariant ensembles of random matrices,
one is able to prove universality in the bulk because of the availability
of explicit formulae for eigenvalue statistics that are amenable to asymptotic analysis.

A priori, universality is more plausible for unitary, orthogonal and symplectic ensembles
because of the invariance properties that are built in: such ensembles are
already ``part of the way there.'' This is not the case, however, for Wigner ensembles.
Universality in the bulk for Wigner matrices is a conjecture par excellence that digs
deep into the structure of random matrices. Numerical experiments provide convincing evidence that it is true.

\bigskip

\textbf{Problem 3. Riemann-Hilbert Problem with non-analytic data.} In many situations one is concerned
with the asymptotic behavior of Riemann-Hilbert problems with exponentially varying data of the form
$e^{in\phi(z)} r(z)$, $n \to \infty$. The Deift-Zhou nonlinear steepest descent method for such problems
requires $\phi(z)$ to be analytic. The analyticity is used in two ways: to control the equilibrium measure
associated with the problem, and then to deform the contour for the RHP. By contrast, the method only requires minimal smoothness for $r(z)$ (see eg. Deift-Zhou in the context of Problem 12 below). It is of considerable
theoretical and practical interest to extend the nonlinear steepest descent method to situations where
$\phi$ is no longer analytic, and has, for example, only a finite number of derivatives. For very interesting
work on the analyticity problem, we refer the reader to a recent paper of Miller and McLaughlin. There is also
interesting, older work due to Varzugin.

\bigskip

\textbf{Problem 4. Painlev\'e equations.} What I have in mind here is not a specific problem, but
a project, a very large scale project. The six (nonlinear) Painlev\'e equations form the core of
``modern special function theory.'' The role that the classical special functions, such as the Airy,
Bessel, and Legendre functions, started to play in the 19$^\text{th}$ century, has now been greatly
expanded by the Painlev\'e functions. Increasingly, as nonlinear
science develops, people are finding that the solutions to an extraordinarily broad
array of scientific problems, from neutron scattering theory, to PDEs, to
transportation problems, to combinatorics,..., can be expressed in terms of
Painlev\'e functions.
What is needed is a project, similar to the Bateman project, or a new volume
of Abramowitz and Stegun, devoted to the Painlev\'e equations.
Much can be, and has been, proved regarding
the algebraic and asymptotic properties of Painlev\'e functions. Here the role of integral representations
and the classical steepest descent method in deriving precise asymptotics and connection formulae
for the classical special functions is played, and expanded, by a Riemann-Hilbert
representation of the Painlev\'e equations, together with the non-commutative steepest
descent method introduced in 1993. Very little is known, however, beyond ad hoc calculations,
about the numerical solution of the Painlev\'e equations. If $u(x)$ is the solution of
the Painlev\'e II equation, say, which is asymptotic to the Airy function $Ai(x)$ as
$x\to+\infty$, one would like to know, for example, the location of its poles in the
complex $x$-plane. A modern ``Bateman Project: Painlev\'e equations'' would not/should not
provide tables for such solutions. Rather, it should provide reliable, easy to use software to compute the solutions.
Writing useful software for such nonlinear equations presents many challenges, conceptual,
philosophical and technical. Without the help of linearity, it is not at all clear how to select a
broad enough class of ``representative problems.'' The software should be in the form of a
living document where new numerical problems can be addressed by a pool of experts as they arise. And
at the technical level, how does one combine asymptotic information about the solutions
obtained from the Riemann-Hilbert problem, together with efficient numerical codes
in order to compute the solution $u(x)$ at finite values of $x$?

I believe that the importance of the Painlev\'e Project will only grow with time. It should
be viewed as creating a national resource and should probably be funded and led at the national level.
The NIST Project ``Digital Library of Mathematical Functions'', where Peter Clarkson
has a contribution on Painlev\'e functions, is an encouraging first step.

\bigskip

\textbf{Problem 5. Multivariate analysis.} Random matrix theory was introduced
into theoretical physics by Wigner in the 1950s in his study of neutron scattering resonances,
but as a subject, RMT goes back to the work of statisticians at the beginning of the 20th century.
Recently, advances in RMT have opened the way to the statistical analysis of data sets in cases where the number of
variables is comparable to the number of samples, and both are large. One might, for example,
be interested in the daily temperature in hundreds of cities around the world, over a
365 day time period. At the technical level, one considers the statistics of the singular values of
(appropriately centered and scaled) $p\times n$ matrices $M=(M_{ij})$, where
$p\sim n\to\infty$. Here $p$ is the number of variables and $n$ is the sample size.
More precisely, one centers the $M_{ij}$'s around their sample averages,
$$
M_{ij}\to\hat M_{ij}=M_{ij}-\frac1n\sum_{k=1}^n M_{ik},
$$
and considers the eigenvalues $l_1\geq\cdots\geq l_p\geq0$ and associated
eigenvectors $w_1,\dots,w_p$ of the $p\times p$ sample matrix
$S=\frac1n \hat M \hat M^T$. The $l_i$'s and $w_i$'s are known as the
principal component eigenvalues and eigenvectors, respectively. In Principal
Component Analysis (PCA) ``significant'' dimension reduction in the data occurs if the first few principal components
$l_1, l_2,...$ account for a ``high'' proportion of the total variance $\tr S=\sum_{j=1}^p l_j$.

A common model for the variables $M_{ij}$ is to assume that they follow a (real) $p$-variate Gaussian
distribution $N_p(\mu,\Sigma)$ with mean $\mu$ and covariance matrix $\Sigma$. Thus the columns
$(M_{1j}, M_{2j},\dots, M_{pj})^T$ provide $n$ independent samples for $N_p(\mu,\Sigma)$. Using recent results from RMT,
much has now been proved about the statistics of $l_1, l_2,...$ as $p,n\to\infty$, $p/n\to\gamma\in(0,\infty)$,
in the case $\Sigma=I$. In particular, we know that in the limit, $l_1$, appropriately centered and scaled,
satisfies the Tracy-Widom distribution for the largest eigenvalues of a GOE matrix. However, most interesting
applications involve so-called spiked populations, a terminology introduced by Johnstone,
i.e. situations where most of the eigenvalues $\eta_1,...,\eta_p$ of $\Sigma$ are equal to
a common value, say 1, but the first few eigenvalues are greater than 1. Thus
$$
\eta_1\geq\eta_2\geq\cdots\geq\eta_k>\eta_{k+1}=\cdots=\eta_p=1
$$
for some fixed $k<<p$. It is a major problem in multivariate analysis
to analyze the statistics of the eigenvalues $l_1, l_2,...$ as $p,n\to\infty$, $p/n\to\gamma\in(0,\infty)$
for such spiked populations. There are very interesting phase transitions in the theory. For example
if $\eta_1>1+\sqrt{\gamma}$, then, as $p\sim n\to\infty$, $l_1$ emerges from the Marchenko-Pastur
continuum $\bigl((\sqrt\gamma-1)^2,(\sqrt\gamma+1)^2\bigr)$, where most of the $l_j$'s tend to accumulate,
and almost surely
$$
l_1\to\eta_1 \cdot \Bigl(1+\frac{\gamma}{\eta_1-1}\Bigr)>(1+\sqrt\gamma)^2\,.
$$
In the spiked, complex case, i.e. when the columns $(M_{1j},M_{2j},...,M_{pj})^T$
are sampled from the complex $p$-variate Gaussian distribution,
much is known about the asymptotic distribution of the $l_j$'s, as $p,n\to\infty$, $p/n\to\gamma\in(0,\infty)$.
By contrast in the real case, apart from a.s. convergence of the $l_i$'s,
very little is known about their asymptotic distributions. In the spiked, complex case
the analysis is enabled by a particular technical tool, the Harish-Chandra-Itzykson-Zuber
formula: unfortunately, no analog of this formula is known in the real case. While there
are relatively few applications of complex spiked populations, knowledge of the asymptotic
distributions of the $l_i$'s for real spiked populations would have immediate applications
to a wide variety of problems in signal processing, genetics and finance.

\bigskip

\textbf{Problem 6. $\beta$-ensembles.} Random point processes corresponding to
$\beta$-ensembles, or, equivalently, log gases at inverse temperature $\beta$, are
defined for arbitrary $\beta>0$. The orthogonal, unitary, and symplectic ensembles
corresponding to $\beta=1, 2,$ or 4, respectively, are now, of course, well understood, but other values
of $\beta$ are also believed to be relevant in applications, for example, in the statistical
description of headway in freeway traffic. For certain rational values of $\beta$, $\beta$-ensembles are
related to Jack polynomials, but for general $\beta$ much less is known. The analysis of $\beta$-ensembles
for general $\beta$ represents an interesting, and increasingly important, challenge.

Recently there have been significant developments in the theory of general $\beta$-ensembles.
As a result of the work of Edelman and Dumitriu, and also others, we now know that for
all $\beta>0$, there exist (tridiagonal) random matrix models whose eigenvalues are distributed according to $\beta$-ensembles.
Furthermore, taking an appropriate scaling limit of these tridiagonal matrix ensembles,
one arrives at the following remarkable fact. Let $B(x)$, $x\geq0$, denote Brownian
motion, and for any $\beta>0$, let $H_\beta$ denote the Schr\"odinger operator
$\frac{d^2}{dx^2}-x-\frac{2}{\sqrt \beta} dB(x)$ acting on $L^2((0,\infty),dx)$
with Dirichlet boundary conditions at $x=0$. Then (Edelman-Sutton, Ramirez-Rider-Virag) for almost all realizations
$B(x)$, $x\geq0$, $H_\beta$ is self-adjoint with discrete spectrum
$\lambda_1(B,\beta)>\lambda_2(B,\beta)>\cdots$ and for each $k$, $\lambda_k(B,\beta)$
has precisely the same distribution as the $k^\text{th}$ largest eigenvalues
of the corresponding $\beta$-ensemble in the standard edge scaling limit. Part of the challenge in analyzing
general $\beta$-ensembles is to use $H_\beta$ to obtain information about these ensembles.
Already the variational characterization of the $\lambda_k$'s has been used to give simple
proofs of bounds on the asymptotics of the distributions
of the $\lambda_k$'s. Question: can one derive the Tracy-Widom formula for
$\lambda_1(B;\beta=2)$, say, directly from $H_\beta$?

At a more conceptual level, the ($\beta$-ensemble $\leftrightarrow H_\beta$) correspondence
brings random matrix theory front and center into the arena and practice of
modern day probability theory.

\bigskip

\textbf{Problem 7. Non-self adjoint spectral problems.} Much is now known, theoretically and also
numerically, about the spectrum of self-adjoint operators. This is in great contrast to
the situation regarding non-self-adjoint operators, where the spectral
theory is far more subtle and numerical schemes must overcome significant, inherent
instabilities. The heart of the difficulties and instabilities lies in the following fact: in
the self-adjoint case $A=A^*$, the resolvent $(A-\lambda)^{-1}$ is bounded by the distance
of $\lambda$ to the spectrum of $A$, but in the non-self-adjoint
case, this is no longer true, as we see already in the $2\times2$ case,
$A=\begin{bmatrix}0 &n\\0&0\end{bmatrix}$, $n\to\infty$. In recent years,
a number of authors (e.g. Trefethan, Davies, ...) have initiated a systematic
approach to non-self-adjoint spectral problems, notions such as pseudospectrum
have come into prominence, and other authors have conducted in depth
studies of particular non-self-adjoint spectrum problems which arise in practice.
For example, in analyzing the semi-classical limit of the focusing NLS equation,
one must analyze the spectrum of the associated AKNS operator $T(h)$,
as Planck's constant $h$ goes to zero. The operator is non-self-adjoint,
and as $h$ goes to zero more and more eigenvalues,
corresponding to solitons, emerge in the complex plane. It is of critical
importance to the analysis of the semi-classical limits for NLS to
determine where in the plane, and at what rate, the eigenvalues accumulate.
The difficulty in doing this, theoretically and numerically,
is illustrated by the following fact: the spectrum of $T(h)$ off the real axis is a discrete set,
whereas the numerical range of $T(h)$ is an open subset of the plane. Nevertheless,
every point $\lambda$ lying in the numerical range of $T(h)$ is an eigenvalue
of $T(h)$ to all orders in $h$, $\|(T(h)-\lambda)u\|=O(|h|^k)$
for any $k\geq 1$, for some $u=u(h)$, $\|u\|=1$. In the language of Kruskal,
the computation of the spectrum of $T(h)$ is a problem ``beyond all orders.''
Much has been done (Kamvissis-Miller-McLaughlin, Tovbis-Venakides-Zhou,...) in analyzing
the semi-classical limit for NLS in special situations where the spectral
problem can be solved explicitly. The spectral problem with general data, both for
NLS and also other related non-self-adjoint problems, however, is far from
understood, and poses a great challenge whose resolution is still only in the initial stages.

\bigskip

\textbf{Problem 8. Long-time behavior with non-generic initial data.} The
long-time behavior of the solution of the Cauchy problem for a great many integrable systems on the line
is now well-understood, using, for example, the Riemann-Hilbert/steepest descent method. The method depends on
full knowledge of the nature of the spectrum of the associated Lax operator. For systems such as
KdV, MKdV, defocusing NLS and the Toda lattice, for example, one is able to describe the solution asymptotically
in complete detail for general initial data. But for focusing NLS, where the associated AKNS
Lax operator $T$ is non-self-adjoint (here we make contact with Problem 7), the
situation is different. For generic $T$ (i.e. an open dense set of $T$'s in any reasonable topology)
the spectrum consists of the real line, where the spectrum is absolutely continuous, together with a finite
number of simple eigenvalues (corresponding to solitons) off the real axis. The analysis of the
long-time behavior of focusing NLS with such generic initial data proceeds in a straightforward
manner similar to KdV, MKdV, etc. For general initial data, however, the situation is more complicated.
For example, let $z_0>0$ be any positive number and let $D$ be any arbitrarily small open
disk in the complex plane centered at $z_0$, such that $D\subset\{z\,:\,\Re z>0\}$. Let
$D^+$ denote the intersection of $D$ with the upper half plane, and let $u(z)$
be an arbitrary function analytic in $D^+$, and continuous in $\overline{D^+}$. Let $B=\{z\in D^+\,:\, u(z)=0\}$.
Then there exists (Zhou) an AKNS operator $T$ with infinitely smooth, rapidly decaying coefficients
with the property that each point in $B$ is an $L^2(\mathbb{R})$-eigenvalue of $T$. In other words, there exist (non-generic)
operators $T$ with Schwartz space coefficients which have $L^2$ spectrum accumulating on the real line at an essentially
arbitrary rate. It is a very interesting question to determine what effect such singularities would have on the
long-time behavior of the solution of NLS.
In particular, recalling that focusing NLS provides a model for data transmission
along communication cables, are the effects measurable?

The difficulty that we encounter here is not limited to situations where the associated Lax operator
is non-self-adjoint. Even in situations when the associated operator is self-adjoint, but of order greater than
two, similar difficulties can arise. This is true, in particular, for the Boussinesq equation, where the associated
Lax operator is third order. The long-time behavior of the solutions of the Boussinesq equation with general initial
data is a very interesting problem with many challenges. Even in the case with generic initial data the situation
is only partially understood.

\bigskip

\textbf{Problem 9. The parking problem.} A number of so-called ``transportation'' problems have
now been analyzed in terms of RMT. These include: the ``vicious'' walker problem of M. Fisher,
the bus problem in Cuernavaca, Mexico, the headway traffic problem on highways, and the airline
boarding problem of Bachmat et al. Recently, researchers in London, Prague, and also Ann Arbor,
have noticed an intriguing phenomenon. They found that the fluctuations in the
spacings between cars parked on a long street exhibited RMT behavior. Furthermore,
\v{S}eba found that there was a difference whether the street is two-way or one-way
(On a two-way street,
the cars park only on the right, while on a one-way street one of course has the option of
also parking on the left.) Quite remarkably, for two-way streets \v{S}eba
found GUE statistics, but for left-side parking on one-way streets he found GOE statistics.
It is a great challenge to develop a microscopic model for the parking problem,
in analogy, perhaps, with the microscopic model introduced by Baik et al. to explain the RMT statistics
for the bus problem in Cuernavaca. \v Seba's recent, intriguing calculations on the parking problem can be found posted on the web.

\bigskip

\textbf{Problem 10. A Tracy-Widom Central Limit Theorem.} The fact that RMT, and the Tracy-Widom distributions,
arise in so many problems in so many different areas leads one to the following question:
how can one characterize RMT in purely probabilistic terms? For example, we know that if we take i.i.d.'s
$(a_1,a_2,...)$, add them up, and then center and scale appropriately,
$$
(a_1,a_2,...)\to(S_1,S_2,...),\qquad S_n=\frac{\sum_{i=1}^n a_i-n\mu}{\sqrt n},
$$
then as $n\to\infty$, $S_n$ converges in distribution to a Gaussian random variable: this
is the famous Central Limit Theorem. The analogous situation for RMT is the following:
take i.i.d.'s $(a_1,a_2,...)$, perform an operation $X$ on them,
$$
(a_1,a_2,...)\to(X_1,X_2,...),
$$
and as $n\to\infty$ the $X_n$'s converge to the Gaudin distribution, or the Tracy-Widom
distribution. The question is, ``What is $X$?'' Important progress towards answering this question
has been made recently, and independently, by Baik-Suidan and Bodineau-Martin, but the full
problem remains open and very challenging.

\bigskip

\textbf{Problem 11. The Toda lattice with random initial data.} A fundamental question in numerical analysis
is the following: how long does it take on average to diagonalize a random symmetric $n\times n$ matrix $M$?
There are different opinions about what is meant by random, and also many eigenvalue algorithms that one could choose.
Also one must specify the accuracy that is required. For definitiveness, let us assume that $M$ is chosen from GOE
and that we use the standard QR eigenvalue algorithm. The fundamental question then takes the following more concrete
form: given $\epsilon>0$, how many QR steps does it take on average to compute the eigenvalues of a GOE random matrix
to order $\epsilon$? In practice, one never computes the eigenvalues of a full $n\times n$ matrix $M$. Rather one first
reduces $M$ to a tridiagonal matrix
$$
J=
\begin{bmatrix}
a_1 & b_1 & & \\
b_1 & a_2 & \ddots &  \\
    &\ddots&\ddots & b_{n-1} \\
    &      & b_{n-1}& a_n
\end{bmatrix}
$$
with the same spectrum as $M$ using, say, a succession of Householder transformations. Now it turns out,
by pure serendipity, that the Householder transformation is eminently compatible with GOE and the statistics of
$J$ can be computed explicitly: the $a_i$'s and $b_j$'s are independent, the $a_i$'s are i.i.d. Gaussians,
and the $b_j$'s have $\chi$ distributions, $\chi_{n-j}$, $1\leq j\leq n-1$. Such matrices
are an example of what we may call a TE1 (triangular ensemble, $\beta=1$). With these comments, the fundamental question
now takes the following sharp form: given $\epsilon>0$, how many QR steps does it take on average to
compute the eigenvalues of a TE1 matrix $J$ to order $\epsilon$?

In another, independent development in the 1970s, Flaschka, and also Manakov, showed that the Toda lattice,
consisting of $n$ 1-dimensional particles interacting with exponential forces
\begin{equation}\label{E:5}
\ddot{x}_k=e^{x_{k-1}-x_k}-e^{x_k-x_{k+1}},\qquad 1\leq k\leq n,
\end{equation}
$(x_0\equiv-\infty, x_{n+1}\equiv+\infty)$ is a completely integrable Hamiltonian system
with a Lax operator given by the tri-diagonal matrix
$$
J=
\begin{bmatrix}
a_1 & b_1 & & \\
b_1 & a_2 & \ddots &  \\
    &\ddots&\ddots & b_{n-1} \\
    &      & b_{n-1}& a_n
\end{bmatrix}\,,
$$
where
$$
\left\{
  \begin{array}{ll}
    a_i=-\frac{1}{2}\dot x_i, & 1\leq i\leq n \\
    b_i=\frac12 e^{\frac12(x_i-x_{i+1})}, & 1\leq i\leq n-1.
  \end{array}
\right.
$$
With these variables, \eqref{E:5} takes the isospectral form
\begin{equation}\label{E:6}
\frac{dJ}{dt}=[B,J]=BJ-JB,
\end{equation}
where
$$
B=
\begin{bmatrix}
0 & b_1 & & \\
-b_1 & 0 & \ddots &  \\
    &\ddots&\ddots & b_{n-1} \\
    &      & -b_{n-1}& 0
\end{bmatrix}.
$$
Subsequently Moser integrated \eqref{E:5} explicitly and showed that
as $t\to\infty$ the particles become free,
\begin{equation*}
\left\{
  \begin{array}{ll}
    \dot x_i(t)=\dot x_i(\infty)+o(1), & 1\leq i\leq n \\
    x_i(t)=t\dot x_i(\infty)+x_i(\infty)+o(1), & 1\leq i\leq n,
  \end{array}
\right.
\end{equation*}
for some constants $\{x_i(\infty)\}_{i=1}^n$ and $\{\dot x_i(\infty)\}_{i=1}^n$,
where $\dot x_n(\infty)>\dot x_{n-1}(\infty)>\cdots>\dot x_1(\infty)$. Said
differently, as $t\to\infty$, $J(t)$ converges to a diagonal matrix
$\text{diag}\left(-\frac{\dot x_1(\infty)}{2},\dots,-\frac{\dot x_n(\infty)}{2}\right)$.
But as $J(t)$ undergoes an iso-spectral evolution, it follows that
$a_i(\infty)=-\frac{\dot x_i(\infty)}{2}$, $1\leq i\leq n$, must be the eigenvalues
of the original matrix $J(t=0)$,
where $a_i(0)\equiv -\frac{\dot x_i(0)}{2}$, $1\leq i\leq n$,
$b_j(0)\equiv\frac12 e^{\frac12(x_j(0)-x_{j+1}(0))}, 1\leq j\leq n-1$.
Deift-Li-Nanda-Tomei then raised the possibility of using the Toda lattice as an eigenvalue algorithm
(the ``Toda algorithm''): given a tridiagonal matrix $J_0$ and $\epsilon>0$, solve \eqref{E:6}
with initial condition $J(t=0)=J_0$, until $\max_{1\leq i\leq n} b_i(t)<\epsilon$. Then
$a_i(t), 1\leq i\leq n$, are the eigenvalues of $J_0$ to order $\epsilon$. Using earlier
work of Symes, Deift et al showed that the QR algorithm itself was the time-$k$
map, $k=1,2,...$, of a completely integrable Hamiltonian system which Poisson commutes with the Toda Hamiltonian. In this picture,
the choice of an eigenvalue algorithm becomes simply the choice of a Hamiltonian vector field. In view
of the above comments and remarks, the fundamental question becomes: how long does it take on average
for Toda particles with TE1 initial data to become free (i.e. $\max_{1\leq i\leq n-1} b_i<\epsilon$)?
Taking into account the ideas of deflation, it is enough to consider the time $t$ at which
just $b_{n-1}(t)<\epsilon$.

The solution of the fundamental question in the above form is a fascinating challenge and would clearly be of
great interest to scientists and engineers alike.

\bigskip

\textbf{Problem 12. Perturbation theory for infinite dimensional integrable systems.} The bijective mapping properties of the
scattering transform for a variety of integrable systems on the line between appropriate weighted Sobolev spaces
have now been established (X.~Zhou) using Riemann-Hilbert techniques. This has made it possible,
in particular, to analyze (Deift-Zhou) the long-time behavior of the solution
of the Cauchy problem for a variety of integrable systems in a fixed space without a ``loss of derivatives''.
This in turn has made it possible to analyze perturbations of integrable systems. For example, Deift-Zhou analyzed the
perturbed defocusing NLS equation
\begin{equation}\label{E:7}
iu_t+u_{xx}-2|u|^2u-\varepsilon V(|u|)u=0
\end{equation}
$$
u(x,t=0)=u_0(x)\in H^{1,1}=\{f\in L^2\,|\, f', xf\in L^2(\mathbb{R})\}\,,
$$
where $V(|u|)\sim |u|^p$ as $|u|\to0$, for some $p>2$ sufficiently large. In addition to a full description
of the long-time behavior of the solution of \eqref{E:7}, a rather surprising outcome of their calculations is
that \eqref{E:7} remains completely integrable for all $\varepsilon$ (small and) positive. In the perturbation theory
of the linear Schr\"odinger equation -- when the term $-2|u|^2u$ is absent from \eqref{E:7} -- the key role
in the analysis is played by the Fourier transform which diagonalizes the linear part of the equation.
For the full equation \eqref{E:7}, the role of the Fourier transform is now played by the scattering transform
which ``diagonalizes'' the (cubic) NLS part of the equation. In the case of the linear Schr\"odinger equation, the method
relies on certain precise estimates that one obtains from the Fourier transform and the classical steepest descent
method; in the fully non-linear case \eqref{E:7}, the analogs of these estimates are obtained from the Riemann-Hilbert
version of the scattering/inverse scattering transform, together with the steepest descent
method for RH problems. One shows, in particular, that $\lim_{t\to\infty}(\sup_{x\in\textbf{R}} |u(x,t)|)=0$,
which implies that as $t\to\infty$ the perturbation term $\varepsilon V(|u|)u\sim \varepsilon |u|^p u$,
$p>2$, becomes small with respect to the NLS term $2|u|^2u$.

It is of great interest to consider perturbations of type \eqref{E:7} for the focusing (cubic) NLS equation,
i.e. where the term $-2|u|^2u$ is replaced by $+2|u|^2u$. This changes the problem fundamentally
as the focusing equation has soliton solutions. Such solutions do not decay uniformly in time
(i.e. $\sup_{t,x\in\mathbb{R}}|u(x,t)|>0$) and this complicates
the analysis of the perturbed equation enormously, as the perturbation term is no longer
small with respect to $2|u|^2u$ (see above). Equation \eqref{E:7} in the focusing case
is just one example of many and varied systems in PDE/Mathematical Physics. In higher dimensions,
the behavior of such perturbed systems in the neighborhood of a simple non-linear bound state
is well understood (Soffer, Weinstein,...), but for more than one bound state, very little is known.

The solution of \eqref{E:7} in the focusing case in the neighborhood of a $k$-soliton for (cubic) NLS ($k\geq2$),
together with a detailed description of the long-time asymptotics, would be regarded as a very significant
development in the theory of PDEs/Mathematical Physics. In a certain sense,
one expects \eqref{E:7} once again to be integrable for small $\varepsilon$. Equation
\eqref{E:7} is of course only one of many similar systems one could consider. For example, one
could consider whether the Toda shock and Toda rarefaction phenomena persist when the exponential forces
of interaction are perturbed: all numerical evidence indicates that this is so. Also, one could consider perturbations
of the sine-Gordon equation as a model for the Fermi-Pasta-Ulam problem.

There is a considerable body of work (Kuksin, Kapeller-P\"oschel) on Hamiltonian perturbations of
integrable systems such as NLS or KdV in the spatially periodic case. Here KAM methods apply
and the authors show that certain finite dimensional tori corresponding to finite gap solutions in the integrable
case survive under perturbation. The periodic problem is more complicated than the problem on the line
because of the action of dispersion. On the whole line dispersion forces soliton-free solutions $u(x,t)$ to
decay, $\sup_{x\in\mathbb{R}} u(x,t) \to 0$ as $t\to\infty$, and even for solutions with solitons (cf. comments above)
we still expect decay away from the location of the solitons. In the periodic case, dispersion has no ``room''
to act and the perturbation term $\varepsilon V(|u|)u$ is of the same order as $|u|^2u$ for all time.

\bigskip

\textbf{Problem 13. Perturbation theory for exactly solvable combinatorial problems.}
The asymptotic behavior of a variety of combinatorial problems has now been analyzed in great detail. Here we have in mind Ulam's
problem for the length of the longest increasing subsequence, the tiling problem for the Aztec diamond, the hexagon tiling problem,
and the last passage percolation problem, amongst many others. In all cases, in an appropriate scaling limit, the statistical
fluctuations in the systems at hand are described by random matrix theory.
The asymptotic analysis, however, depends in a critical, and rigid way, on the underlying probability measures for the systems.
For example, the analysis of the last passage percolation problem requires the waiting time $w_i$ at each site
$i$ to be either geometrically or exponentially distributed (Johansson): if the statistics of the waiting times is neither
geometric nor exponential, the analysis fails completely. What happens if the geometric distribution, say, is slightly perturbed?
The challenge here is to develop an effective perturbation theory for such systems. One expects that the random matrix behavior
of the fluctuations should persist. Related work on this problem has been done by Baik-Suidan and Bouchard-Martin (cf. Problem 10).

\bigskip

\textbf{Problem 14. Initial/boundary value problems for integrable systems.}
In the late 1990s Fokas introduced a new, more flexible approach to inverse scattering theory, and in
recent years a number of researchers (Fokas, Its, ..., Anne Boutet de Monvel, Shepelsky,...) have applied Fokas'
approach to the initial/boundary value problem for various integrable systems such as NLS. Much progress has been
made, but there is a basic, and puzzling, obstacle to applying the method, viz. one needs to know certain
dependent data in order to proceed. For example, the initial/boundary value problem for NLS
is well-posed if one gives the initial data $u(x,0)=u_0(x)$, $x\geq 0$, and the boundary data $u(0,t)=u_1(t)$,
$t\geq 0$. However, in implementing the method it turns out that one needs to know the \textit{dependent} data $u_x(0,t)$, $t\geq 0$,
as such information appears explicitly in the solution formulae. Progress has been made in
determining $u_x(0,t)$, $t\geq0$, from $u_0(x)$ and $u_1(t)$ intrinsically via the method, but the
control one obtains on $u_x(0,t)$ is not sufficient in order to obtain the long-time behavior
of the initial/boundary value problem. This is true even in simple cases, such as the following:
suppose $u(x,t)$ solves NLS in $(x\geq0,t\geq0)$ with $u(0,t)=\sin(\omega t)$, $\omega\neq0$, and
$u(x,0)=u_0(x)$, where $u_0$ is smooth with compact support. How does $u(x,t)$ behave
as $t\to\infty$? The solution of this problem for any such $u_0(x)$ would be a very significant
development in the theory, and would also be of considerable
interest in science and engineering.

There is a philosophical point at stake here. The evolution of NLS in $x>0$ represents the interplay
of forces which are ``integrable'' at some fundamental, algebraic level, and indeed, if no other forces are present,
as in the full-line scattering or periodic cases, the equation can be integrated explicitly. When one considers
the initial/boundary value problem, however, new physical forces come into play which describe the interaction
of the particles on the boundary with the NLS particles in the interior. There is absolutely no
a priori guarantee that the enlarged system, ``NLS particles in $x>0$''+``particles on the boundary'', is integrable.
It may be that the long-time behavior of the composite system can only be solved for
``generic, Cantor-set'' like data, as is familiar from KAM theory. In other words, an explicit description of
the long-time behavior of the solution of the initial/boundary value problem for NLS
for general initial data may not be possible. This is a very intriguing situation.

\bigskip

\textbf{Problem 15. Multi-matrix models and models with an external field.} There has been considerable progress
(Kuijlaars,...) in understanding basic statistics such as the correlation functions for the
2-matrix random matrix model, and also matrix models with a source. The key element in these
developments has been the successful extension by Kuijlaars et al of the Riemann-Hilbert/steepest
descent method to $3\times3$ Riemann-Hilbert problems. So far only the simplest situations
have been considered. In order to consider the generic situation, one must, in particular, extend the
Riemann-Hilbert/steepest descent method to $n\times n$ Riemann-Hilbert problems. This is
a challenging problem which would have important implications, not only for random matrix models,
but also for problems in other areas, such as Pad\'e-Hermite approximations and
irrationality questions for distinguished real numbers, and multi-orthogonal polynomials.

\bigskip

\textbf{Problem 16. Poisson/Gaudin-Mehta transition.} On the appropriate scale, the bulk eigenvalues
of a random GOE matrix $M$ exhibit Gaudin-Mehta statistics. As noted in Problem 2, the same is believed to be true
for general Wigner matrices. On the other hand, if $M=M^T$ has i.i.d. entries and bandwidth $W=1$ (i.e. $M$ is tridiagonal),
then, on the appropriate scale, the bulk eigenvalues of $M$ exhibit Poisson statistics. As the bandwidth $W$
increases from 1 to $N-1=\text{dim}(M)-1$, the eigenvalue statistics must change from Poisson to Gaudin-Mehta.
A back of the envelope calculation suggests that there should be a (sharp) transition in a narrow
region around $W\approx\sqrt{N}$. This is a well-known, outstanding, open conjecture with many implications
for theoretical physics, particularly wave propagation in random media. There is a substantial
body of work on this problem, which can be easily traced on the web.

\bigskip

There are many other areas, closely related to the problems in the above list, where much
progress has been made in recent years, and where much remains to be done. These include:
totally asymmetric simple exclusion processes (TASEP) following the work of Borodin,
Sasamoto, Ferrari,..., asymmetric simple exclusion processes (ASEP) following the work
of Tracy, Widom,..., orthogonal rational functions (Vartanian, Zhou, McLaughlin), partition
functions for random matrix models and their connections to mappings on Riemann surfaces
(Itzikson, Zuber,..., McLaughlin, Ercolani,..., Guionnet), the representation theory
of ``large'' groups such as the infinite symmetric group or the infinite
unitary group (Olshanski, Borodin, Okounkov,...), and many others. But as I said in the
beginning, my list is not complete or definitive.

\end{document}